\documentclass[a4paper,11pt]{article}

\usepackage{contribution}

\newcommand{\weblink}[2][]{%
    \ifthenelse{\equal{#1}{}}%
    {\textnormal{\url{#2}}}%
    {\textnormal{\href{#2}{#1}}}%
}

\newcommand{\acknowledgements}[1]{%
  \bigskip\bigskip
  \textsf{\textbf{\Large Acknowledgements}} \\[2ex]
  {#1}
  \bigskip
}

\def\beq{\begin{equation}}
\def\eeq#1{\label{#1}\end{equation}}
\def\eeqn{\end{equation}}

\def\beqa{\begin{eqnarray}}
\def\eeqa#1{\label{#1}\end{eqnarray}}
\def\eeqan{\end{eqnarray}}

\let\bar=\overbar

\def\etal{{\it et al.}}

\def\VEV#1{\left\langle{ #1} \right\rangle}

\def\Dslash{\not{\hbox{\kern-4pt $D$}}}
\def\dslash{\not{\hbox{\kern-2pt $\del$}}}

\def\msb{{\bar{\ssstyle M \kern -1pt S}}}

\newcommand{\contribution}[7][]{%
  \clearpage
  \thispagestyle{plain}
  \ifthenelse{\equal{#1}{}}
  {\hypersetup{pdftitle={#2}}}
  {\hypersetup{pdftitle={#1}}}
  \hypersetup{pdfauthor={{#3} {#4}}}
  {\centering\normalfont\LARGE\bfseries\sffamily #2 \par\nobreak}
  \lhead{}
  \chead{%
    \textit{\footnotesize XIV International Conference on Hadron Spectroscopy
      (\weblink[\textit{hadron2011}]{http://www.hadron2011.de}), 13-17 June 2011, Munich, Germany}%
  }
  \rhead{}
  \bigskip
  \begin{center}
    {#3} {#4}\ifthenelse{\equal{#6}{}}{}{\footnote{\weblink[#6]{mailto:#6}}}
    \ifthenelse{\equal{#7}{}}{}{#7} \\
    \textit{#5}
  \end{center}
  \bigskip
}

\renewcommand{\abstract}[1]{%
  \begin{center}
    \begin{minipage}{0.85\textwidth}
      \begin{footnotesize}
        #1
      \end{footnotesize}
    \end{minipage}
  \end{center}
  \bigskip
}

\begin{document}

%
%
%
%
%
{  

\def\pbar{\overline{\mbox{p}}}
\def\pbarp{\overline{\mbox{p}}\mbox{p}}
\def\pbarA{\overline{\mbox{p}}\mbox{A}}
\def\JPsi{J/\Psi}
\def\rhoL{\VEV{\int \rho\cdot dl}}
\def\sform{\VEV{\sigma_{\mbox{\scriptsize form}}}}
\def\sdiss{\sigma_{\JPsi \mbox{\scriptsize N}}}
\def\sesc{\VEV{\sigma_{esc}}}
\def\Nesc{\mbox{N}_{\mbox{\scriptsize esc}}}
\def\Nform{\mbox{N}_{\mbox{\scriptsize form}}}
\def\Lint{\int L\; dt}
\def\Pfermi{\mbox{P}_{\mbox{\scriptsize F}}}
\def\Ecm{\mbox{E}_{\mbox{\scriptsize cm}}}
\def\totwo{^{\mbox{\scriptsize 2}}}

%

\contribution[$\JPsi$-Nucleon dissociation cross section]  
{Measuring the $\JPsi$-Nucleon dissociation cross section with PANDA}  
{Paul}{B\"uhler}  
{Stephan Meyer Institute for Subatomic Physics \\
 Austrian Academy of Sciences \\
 Boltzmanngasse 3 \\
 AT - 1090 Vienna, Austria}  
{paul.buehler@oeaw.ac.at}  
{}  
%

\abstract{%

With the PANDA detector at the HESR at FAIR it will be possible to study the
production and absorption of charmed hadrons in nuclear targets. Of special
interest in this context is the determination of the $\JPsi$-nucleon
dissociation cross section. This can be determined with measurements of the
$\JPsi$ yield in $\pbarA$ reactions using different target materials. The
experiment is described and numerical simulations are presented.}

%

\section{Introduction}

The inelastic $\JPsi$-nucleon cross section $\sdiss$ is important in
understanding the role of the formation of a Quark-Gluon-Plasma, QGP in the
$\JPsi$ suppression observed in high energy nuclear collisions. For an
interpretation of these data the quantitative understanding of the nuclear
effects - those which are not related to QGP formation - that also affect
$\JPsi$ production in nucleus-nucleus reactions is crucial. The inelastic
scattering of the $\JPsi$ state in cold nuclear matter is expected to be the
dominant contribution of this kind. Its strength is described by the
$\JPsi$-nucleon inelastic cross section, $\sdiss$.

$\sdiss$ can be determined with reactions, in which $\JPsi$ are produced in
nuclear medium. It is from the comparison of the expected number of produced
$\JPsi$ and the actually measured number of $\JPsi$ escaping from the nuclear
medium that the dissociation cross section is derived. 

Current experimental values of $\sdiss$ have mainly been obtained from inclusive
hadro- and leptoproduction of $\JPsi$ on nuclear targets \cite{arleo08}. In
these experiments the determination of $\sdiss$ is hampered by co-acting effects
(co-movers, feed down, ...) which affect the number of escaping $\JPsi$s. Thus
the various contributions have to be disentangled to extract the effect of the
$\JPsi$-Nucleon dissociation on the observed suppression of the $\JPsi$ yield.

In $\pbarA$ reactions the momentum of the incident $\pbar$ can be tuned such
that $\JPsi$s are produced exclusively at a relatively well defined momentum.
This will facilitate the analysis of the data significantly.

\section{The experiment}

The idea for this experiment for measuring $\sdiss$ can be traced back to a
paper by Brodsky \& Mueller \cite{brodsky88}. The principle of the measurement
is illustrated in figure \ref{fig01}a). $\JPsi$s are formed in $\pbarA$
reactions, when an incident $\overline{p}$ annihilates with a nucleon within a
target nucleus. The momentum of the incident antiproton is tuned such that the
center-of-mass energy $\Ecm$ of the antiproton-nucleon system is close to the
resonance energy of the $\JPsi$ of $3.097$ GeV/c$\totwo$ and $\JPsi$ is formed
exclusively. The formed $\JPsi$ has an initial velocity and sets off
to escape from the nucleus. There are two possible fates for the $\JPsi$ -
either it will decay due to its finite lifetime or it will be absorbed in an
inelastic reaction with a nucleon.

The number of $\JPsi$ formed in the nucleus $\Nform$ and the number of $\JPsi$
escaping the Nucleus $\Nesc$ are related through

\beq
\Nesc = \Nform \left ( 1 - \sdiss\rhoL \right )
\eeq{eq01}

$\rhoL$ is the average integrated nucleon density along the path of the
$\JPsi$ through the nucleus, where the average is to be taken over all
formations.

Equation (\ref{eq01}) represents a recipe for the measurement of $\sdiss$. It
contains a few quantities which either need to be determined experimentally
($\Nesc$) or by simulation/calculation ($\Nform$, $\rhoL$). The success of the
measurement thus depends on the exact determination of the absolute values of
these quantities.

\begin{figure}[h]
  \begin{center}
    a) \includegraphics[width=0.45\textwidth]{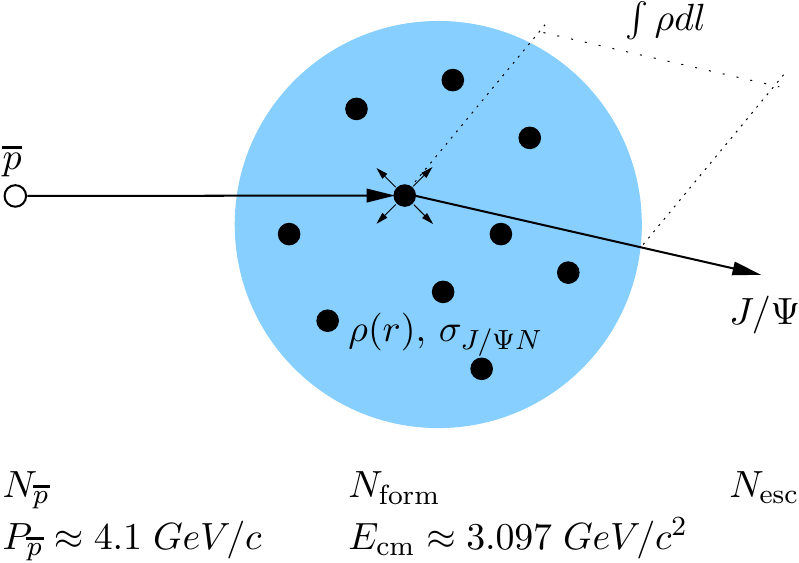}\hfill%
    b) \includegraphics[width=0.45\textwidth]{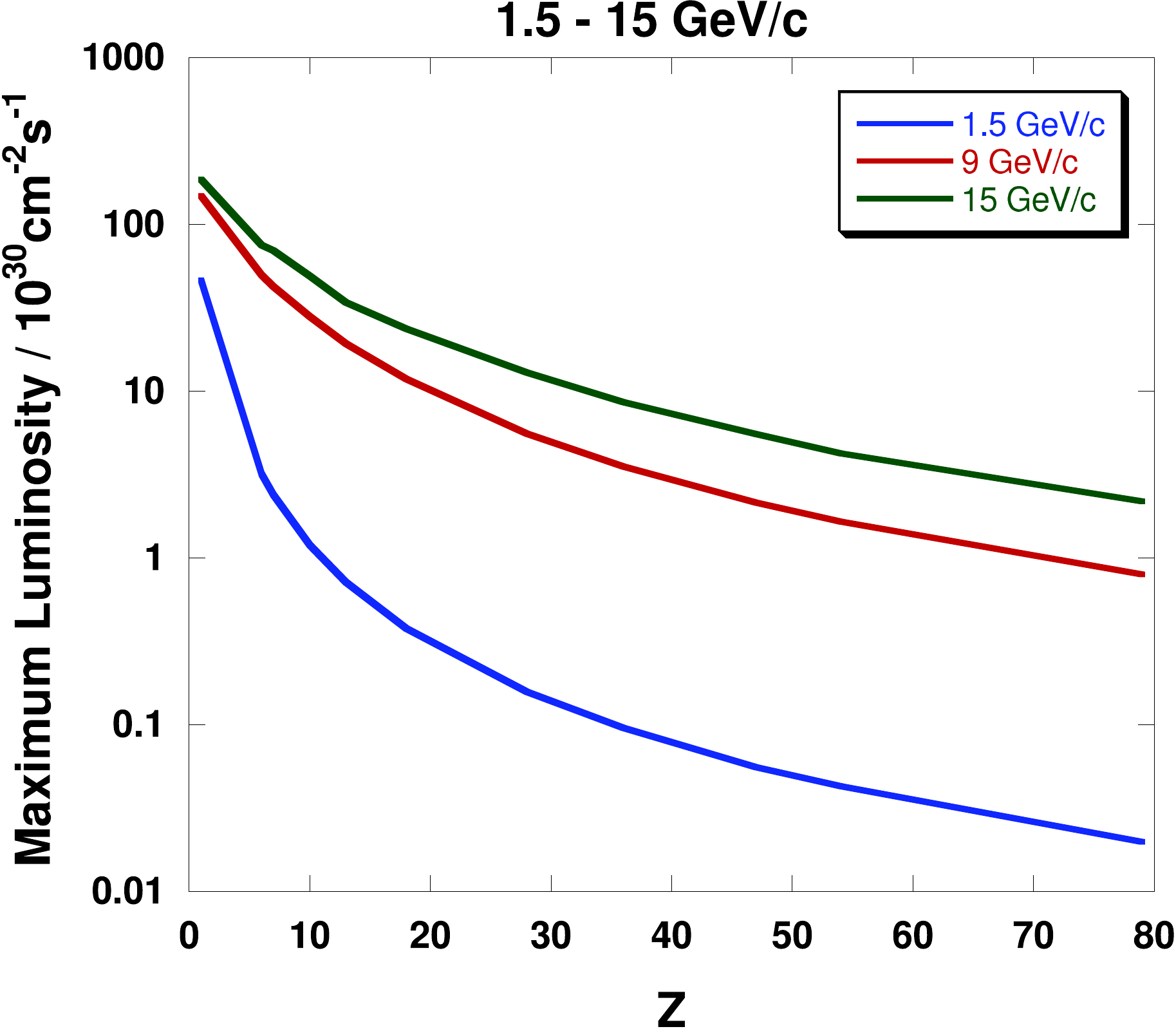}
		\caption{a) Illustration of the $\overline{p}A$ reaction which is used to
		measure the $J/\Psi$-Nucleon dissociation cross section $\sdiss$. b)
		Luminosity of the HESR with a nuclear target as function of the atomic
		number of the target material \cite{ppp09}.}
    \label{fig01}
  \end{center}
\end{figure}

\section{Measurement with PANDA}

$J/\Psi$s are experimentally best identified by their decay into lepton pairs
$e^+e^-$/$\mu^+\mu^-$. The corresponding branching ratio is $5.9~\%$ for either
of these two decay channels \cite{rpp2010}. The simple topology in this reaction
allows to efficiently identify $J/\Psi$s decaying into lepton pairs with a
detector like PANDA and also can be exploited to reduce the background to an
acceptable level. The cross section for formation of $\JPsi$ in $\pbarA$ is by a
factor of a few times $10^{9}$ smaller than the total inelastic cross section of
typically $1$ b. Thus a background suppression of at least $10^{-10}$ has to be
and can be achieved with PANDA (see \cite{ppp09} for further details). The
reconstruction efficiency for the decay channel $\JPsi\rightarrow e^+e^- $ was
estimated to be around $70\%$ and somewhat less for $\JPsi\rightarrow
\mu^+\mu^-$. 

The number of formed $J/\Psi$, $\Nform$ is given by

\beq
\Nform = \Lint \cdot \sform
\eeq{eq02}

$\Lint$ is the luminosity integrated over the measurement time. $\sform$ is the
average formation cross section.

The measurement of the luminosity will have to be carried out instantaneously
with a dedicated luminosity monitor. The achievable luminosity at the High
Energy Storage Ring, HESR at the future FAIR with a nuclear target is shown in
figure \ref{fig01}b) \cite{ppp09}. Due to the enhanced absorption and scattering
in heavier targets, the luminosity decreases with increasing atomic number of
the target material. At the $J/\Psi$ resonance momentum the achievable
luminosity will range from $\approx 5\;\frac{pb^{-1}}{d}$ at $Z\le10$ to
$\approx 10^{-2}\;\frac{pb^{-1}}{d}$ at $Z\ge 40$.

The formation probability in a single $\pbarp$ is well described by the
Breit-Wigner formula and is a function of the available center-of-mass energy
$\Ecm$ (corrections for initial-state radiation can be applied
\cite{kennedy92}). The formation cross section at resonance in a $\pbarp$
reaction and final decay into a $e^+e^-$ pair is $275.7$ nb (relevant
parameters have been determined by \cite{armstrong93}). The effective formation
cross section $\sform$ in $\overline{p}A$ reactions however is considerably
smaller than this peak value. Due to the Fermi motion of the nucleons in the
nucleus, the available $\Ecm$ is determined not only by the energy of the beam
particle but also depends on the Fermi momentum of the involved nucleon. Thus 
$\sform$ is the formation probability averaged over all particular formations.

The value of $\sform$ was determined by Monte Carlo simulations using a Glauber
type model. The nucleus is described by a radial nucleon density profile
$\rho(r)$ which is taken from \cite{devries87,reuter82}. The distribution of the
Fermi-motion is modeled using a local Thomas-Fermi approximation in which the
Fermi momentum $\Pfermi$ depends on the local nucleon density $\rho(r)$ as
$\Pfermi = \hbar c(3\pi\totwo\rho(r))^{\mbox{\scriptsize 1/3}}$. It is assumed, as has been pointed out
by Farrar et al. \cite{farrar90} that the formation time of $J/\Psi$ is short
enough such that color transparency effects can be neglected. 

In this model the formation points are not homogeneously distributed in the
nucleus. The number of antiprotons decreases with increasing depth $s$ due to
the absorption in the nuclear matter. With the total $\overline{p}p$ cross
section of $\approx 7$ fm$\totwo$ at $4$ GeV/c \cite{flaminio79} and a typical
nuclear density of $0.1$ fm$^{\mbox{\scriptsize -3}}$ the penetration depth is
typically in the order of a few fm only. For the Monte Carlo simulation thus
first a location within the nucleus is selected according to the survival
probability of the antiprotons ($\propto e^{-\sigma_{tot}\int_{-\infty}^s
\rho(x,b)dx}$). Then the momentum of the nucleon is selected according to the
local distribution of the Fermi-momentum. Together with the momentum of the beam
particle, which is distributed according to a Gaussian function with a relative
width of typically $10^{\mbox{\scriptsize -4}}$ (this corresponds to the High
Intensity mode of the HESR), and under consideration of the nuclear binding
energy, the $\Ecm$ is computed. This is finally inserted into the Breit-Wigner
formula to compute the formation probability. In order to compute the
dissociation probability the nucleon density along the path of the $\JPsi$ is
integrated up to the point where the $\JPsi$ decays.  Figure \ref{fig02} shows
results of these simulations.

\begin{figure}[htb]
  \begin{center}
    a) \includegraphics[width=0.47\textwidth]{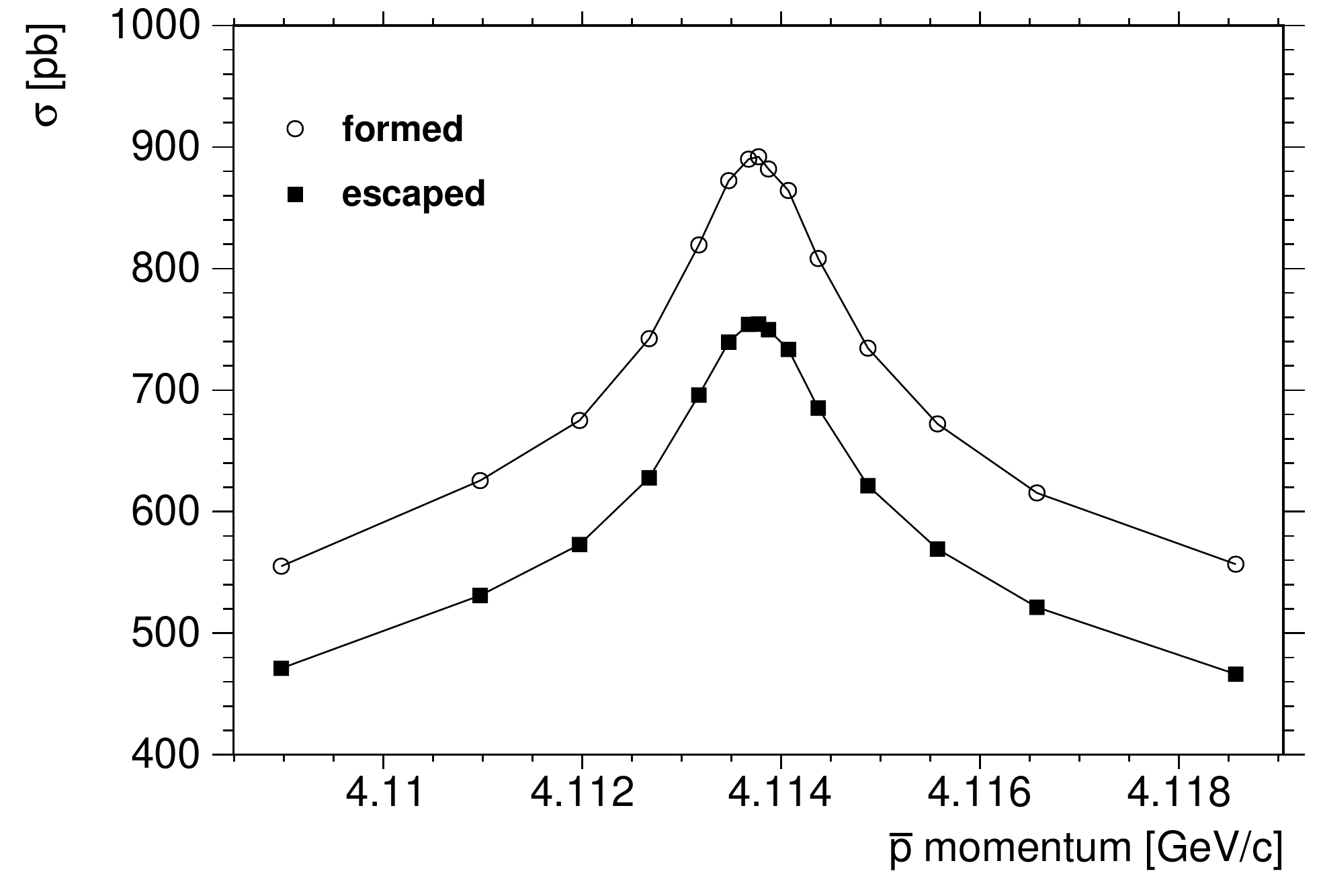}\hfill%
		b) \includegraphics[width=0.47\textwidth]{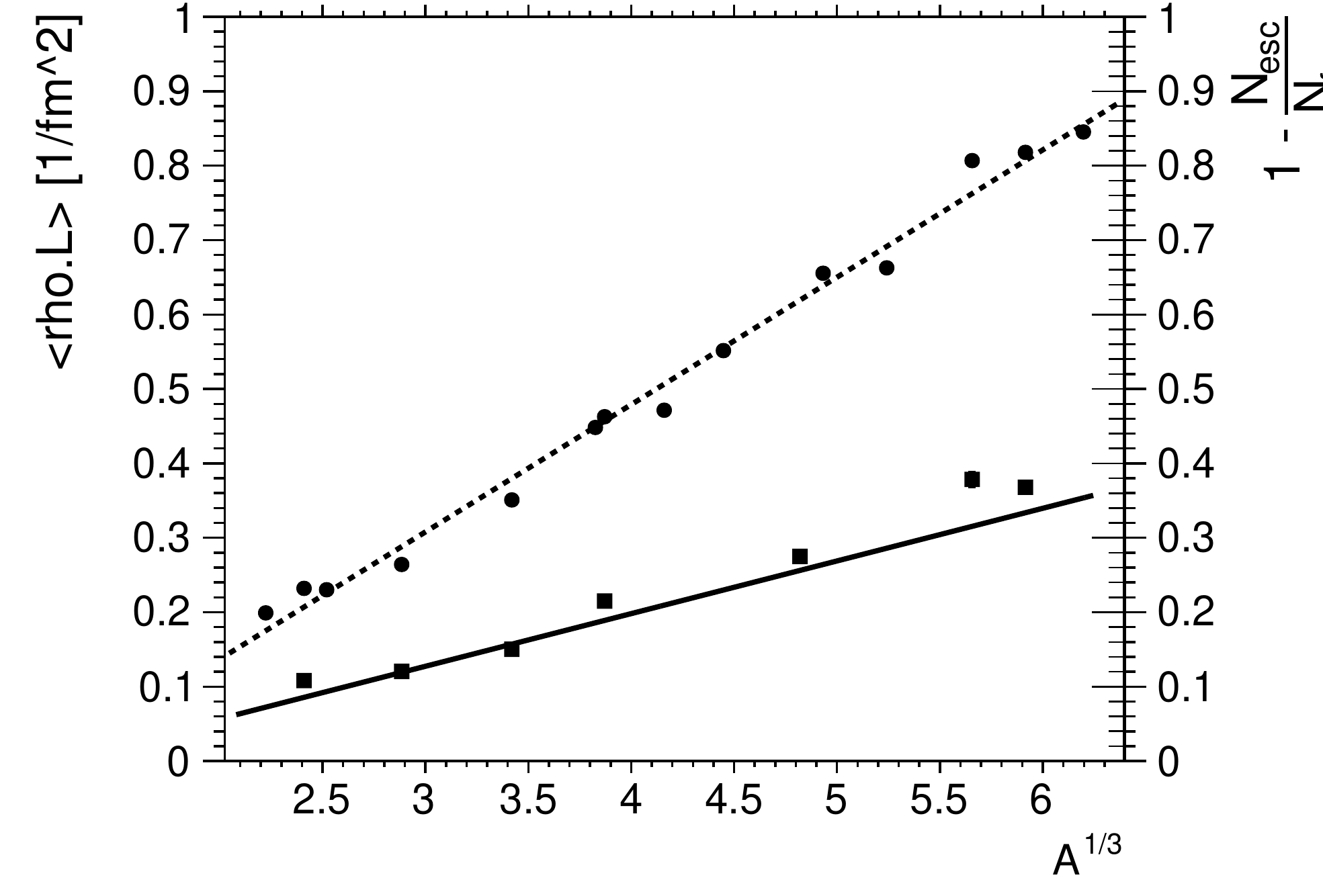} \caption{a) $\sform$
		and $\sesc$ as function of the $\pbar$ momentum in a $^{40}$Ca target b) $(
		1 - \Nesc/\Nform )$ (square symbols) and $\rhoL$ as function of nuclear mass
		of the target material.}
    \label{fig02}
  \end{center}
\end{figure}

In the left panel the expected cross section for formation $\sform$ and the
cross section for escaping and decaying into an electron-positron pair $\sesc$
is plotted as function of the beam momentum. The peak value of $\sesc$ is in the
order of $700$ pb in this case. With a total reconstruction efficiency of
$50\%$ and an average luminosity of $10^{31}$ cm$^{\mbox{\scriptsize
-2}}$s$^{\mbox{\scriptsize -1}}$ the number of $\JPsi$ per day would be around $300$.

According to equation (\ref{eq01}) $\sdiss$ is obtained by the quotient of $( 1
- \Nesc/\Nform)$ and $\rhoL$. For consistency checks, measurements of $\Nesc$
should be carried out for several beam momenta which allow to reconstruct the
broadened resonance and should also be repeated with different target materials.

The square symbols plotted in the right panel of figure \ref{fig02} represent
the quantity $( 1 - \Nesc/\Nform )$ and the circular symbols show the quantity
$\rhoL$ for different target materials. The dotted line is the best linear fit
to the $\rhoL$ points, whereas the bold line is the dotted line scaled to match
the $( 1 - \Nesc/\Nform )$ data points best. The scaling factor is the
measurement of $\sdiss$ and in these simulations its value is found to be very
close to the value which was used as input.

\section{Conclusions}

Measurements of the $\JPsi$ yield in $\pbarA$ reactions close to resonance in
different target materials allow to determine the $\JPsi$-nucleon dissociation
cross section. The experiment depends on an accurate measurement of the number
of $\JPsi$ escaping the nucleus and an accurate prediction of the number of
$\JPsi$ formed in the nuclear medium. The presented simulations suggest, that in
light target materials with this reaction up to a few hundred $\JPsi$ will be
formed and detected with the PANDA detector per day. Measurements on different
target materials will allow to check results for consistency.

\acknowledgements{%
I would like to thank Albert Gillitzer for helpful comments and discussions.}


%

}  


\end{document}